# A multi-dimensional framework for characterizing the citation impact of scientific publications

Yi Bu[1,2], Ludo Waltman[3], and Yong Huang[4]

[1] Department of Information Management, Peking University, Beijing, China
buyi@pku.edu.cn - https://orcid.org/0000-0003-2549-4580

[2] Center for Complex Networks and Systems Research, Luddy School of Informatics, Computing, and Engineering, Indiana University, Bloomington, IN, USA

[3] Centre for Science and Technology Studies, Leiden University, Leiden, The Netherlands
waltmanlr@cwts.leidenuniv.nl - https://orcid.org/0000-0001-8249-1752

[4] School of Information Management, Wuhan University, Wuhan, Hubei, China
yonghuang1991@whu.edu.cn - https://orcid.org/0000-0001-5953-6908

The citation impact of a scientific publication is usually seen as a one-dimensional concept. We introduce a multi-dimensional framework for characterizing the citation impact of a publication. In addition to the level of citation impact, quantified by the number of citations received by a publication, we also conceptualize and operationalize the depth and breadth and the dependence and independence of the citation impact of a publication. The proposed framework distinguishes between publications that have a deep citation impact, typically in a relatively narrow research area, and publications that have a broad citation impact, probably covering a wider area of research. It also makes a distinction between publications that are strongly dependent on earlier work and publications that make a more independent scientific contribution. We use our multi-dimensional citation impact framework to report basic descriptive statistics on the citation impact of highly cited publications in all scientific disciplines. In addition, we present a detailed case study focusing on the field of scientometrics. The proposed citation impact framework provides a more in-depth understanding of the citation impact of a publication than a traditional one-dimensional perspective.

# 1. Introduction

Measuring the citation impact of scientific publications is an important topic in bibliometric and scientometric research. Many different citation impact indicators, calculated based on the citations received by a publication, have been proposed (Waltman, 2016), ranging from the raw citation count of a publication to field-normalized indicators (e.g., Radicchi, Fortunato, & Castellano, 2008; Waltman & Van Eck, 2019; Waltman, Van Eck, Van Leeuwen, Visser, & Van Raan, 2011), recursive PageRank-inspired indicators (e.g., Chen, Xie, Maslov, & Redner, 2007; Walker, Xie, Yan, & Maslov, 2007; Waltman & Yan, 2014) as well as indicators that take into account attributes derived from the full text of citing publications (e.g., Ding, Liu, Guo, & Cronin, 2013; Wan & Liu, 2014; Zhu, Turney, Lemire, & Vellino, 2015). These approaches have in common that they all regard the citation impact of a publication as a one-dimensional concept. In this paper, we propose a multidimensional perspective on the citation impact of a publication. We argue that, in addition to the level of citation impact, there are other relevant aspects of the citation impact of a publication that can be derived from a citation network.

To illustrate this point, consider two publications, $A$ and $B$. As shown in Figure 1, these publications have each received five citations. If we just count the citations received by $A$ and $B$, the publications have the same citation impact. However, the publications citing $A$ also cite each other and therefore seem to be closely related, while the publications citing $B$ do not cite each other and therefore seem to be quite unrelated from each other. Hence, $A$ and $B$ have the same level of citation impact, but they differ fundamentally in the way in which they have an impact on other publications. We say that publication $A$ has a deep citation impact because the publications by which it is cited also cite each other, suggesting that these publications all belong to a relatively narrow research area in which they build on each other in a cumulative way. In contrast, we say that publication $B$ has a broad citation impact because it is cited by publications that do not cite each other. Since the publications citing $B$ do not cite each other, they do not seem to build on each other and they may cover a relatively wide research area. To capture the difference in citation impact between $A$ and $B$, we propose an approach for quantifying the depth and breadth of the citation impact of a publication.

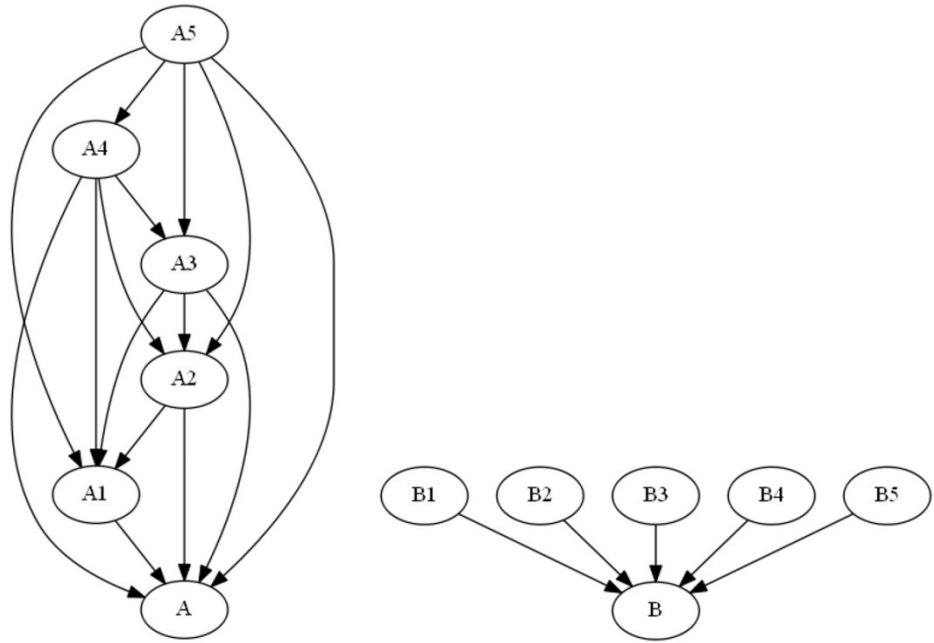

Figure 1. Deep and broad citation impact. Nodes represent publications and edges represent citation relations. Publications *A* and *B* (i.e., the focal publications) have both received five citations. All publications citing *A* (i.e., A1, A2, A3, A4, and A5) also cite each other, while publications citing *B* (i.e., B1, B2, B3, B4, and B5) do not cite each other. Therefore *A* has a deep citation impact, while *B* has a broad citation impact.

We are also interested in the dependence of a publication's citation impact on earlier publications. In Figure 2, publications *A* and *B* have both received five citations, and they both have three references. All publications citing *A* also cite each of *A*'s references, while the publications citing *B* do not cite *B*'s references. Hence, the citation impact of *A* seems to depend strongly on earlier publications, namely the ones cited by *A*. It is likely that *A* is a follow-up study of these earlier publications. In contrast, *B* seems to have a much more independent citation impact, since publications citing *B* do not cite the references of *B*.

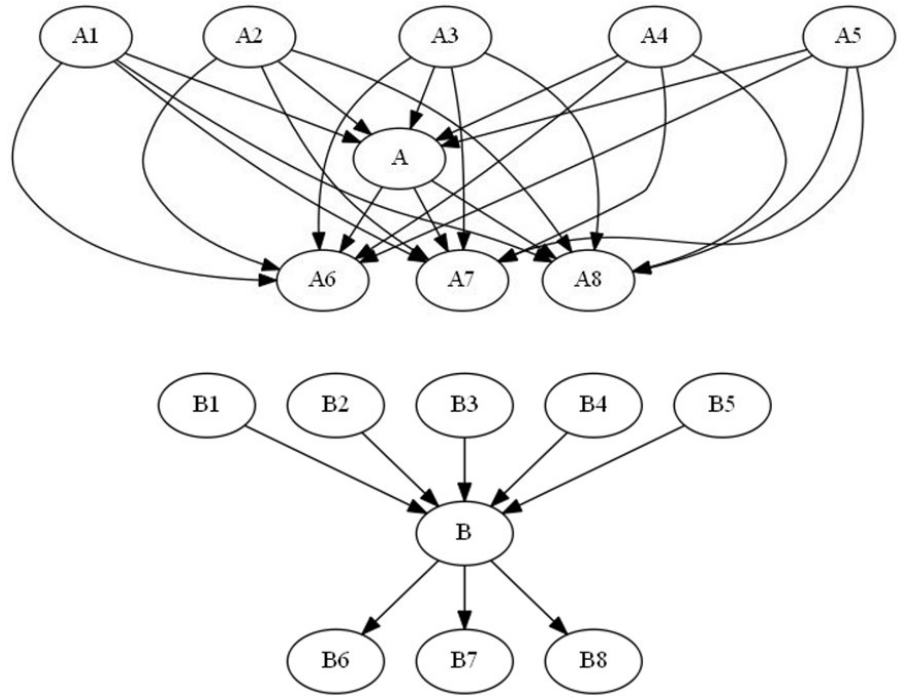

Figure 2. Dependent and independent citation impact. Nodes represent publications and edges represent citation relations. Publications *A* and *B* (i.e., the focal publications) have both received five citations, and they both have three references. All publications citing *A* (i.e., A1, A2, A3, A4, and A5) also cite each of the references of *A* (i.e., A6, A7, and A8), while publications citing *B* (i.e., B1, B2, B3, B4, and B5) do not cite references of *B* (i.e., B6, B7, and B8). Therefore *A* has a citation impact that is strongly dependent on earlier publications, while *B* has an independent citation impact.

We propose to conceptualize and operationalize the citation impact of a publication in a multi-dimensional framework that focuses on (1) the level, (2) the depth and breadth, and (3) the dependence and independence of citation impact. In a traditional one-dimensional perspective on citation impact, only the level of citation impact is considered. Beyond the level of citation impact, no insights are obtained into the way in which a publication has an impact on other publications. By introducing the dimensions of depth and breadth and of dependence and independence, our proposed framework aims to offer a more in-depth understanding of the citation impact of a publication. Our focus is on citation impact at the level of individual publications, but the insights that are obtained at this level may also be used at aggregate levels, for instance at the level of researchers.

The organization of this paper is as follows. In Section 2, we provide a brief discussion of related research. In Section 3, we describe the data that we use in our empirical analyses. In Sections 4-6, we introduce the conceptualization and operationalization of the different dimensions of our citation impact framework (i.e., level, depth and breadth, and dependence and independence) and we report some basic descriptive statistics for each of these dimensions. In Section 7, we present a case study in which we apply our citation impact framework to publications in the field of scientometrics. Finally, in Section 8, we provide some further discussion and we summarize our conclusions.

## 2. Related research

The idea of analyzing citation relations between publications that cite a focal publication has been explored in a number of earlier studies. Clough, Gollings, Loach, and Evans (2015) compared the number of citations given to a publication in a citation network with the number of citations given to the same publication in the transitive reduction of the citation network. According to Clough et al., the transitive reduction can be used to get 'an indication that results in a paper were used across a wide number of fields'. Huang, Bu, Ding, and Lu (2018, 2020) analyzed so-called citing cascades, defined as the citation network of a focal publication and its citing publications. In particular, they studied citation relations between citing publications. The citation impact framework proposed in the current paper partly builds on the ideas explored by Huang et al.

The notion of dependence introduced in our citation impact framework is closely related to the concepts of development and disruption proposed by Funk and Owen-Smith (2017) and used by Wu, Wang, and Evans (2019). Wu et al. investigated an indicator that provides a proxy of whether a publication tends to 'disrupt' or 'develop' science by taking into consideration the publication's references and its citing publications, as well as the citations between all these publications. For a given focal publication, they defined 'type i' publications as those that cite the focal publication but not the references of the focal publication, 'type j' publications as those that cite both the focal publication and the references of the focal publication, and 'type k' publications as those that cite the references of the focal publication but not the focal publication itself. Based on content-level validation, expert interviews, and some other evaluations, the indicator adopted by Wu et al. showed a good performance in assessing the degree to which a publication 'disrupts' or 'develops' science. Yet, a follow-up study by Bornmann and Tekles (2019a) suggested that the length of the time window for calculating the disruptiveness of publications may affect the results. In addition, a case study by Bornmann and Tekles (2019b) questioned the ability of the disruptiveness indicator to identify disruptive publications in the journal *Scientometrics*. Furthermore, Bornmann, Devarakonda, Tekles, and Chacko (2020a, 2020b) compared the disruptiveness indicator with other related indicators, in particular those proposed by Wu and Yan (2019) and by an earlier version of the current paper. Bornmann et al. (2020a) argued that different indicators tend to represent similar dimensions.

The notion of dependence introduced in the current paper is also related to the idea of originality proposed by Shibayama and Wang (2020). Both approaches consider the number of citations from the citing publications of a focal publication to its references.

Building on the idea of citing cascades proposed by Huang et al. (2018), Mohapatra, Maiti, Bhatia, and Chakraborty (2019) introduced a method for pruning the citing cascade of a focal publication $p$. For each citing publication $q$ of $p$, only the longest path between $q$ and $p$ is retained in the pruned network. Based on the pruned network, Mohapatra et al. defined several indicators, in particular depth (i.e., the length of the longest path between the focal publication and the leaf nodes in the network) and width (i.e., the maximum number of nodes at a given level in the network). They assumed that a publication has the most 'influential' impact when the values of depth and width are equal. This assumption lacks a clear conceptual foundation, but it was tested empirically using 'Test of Time Awards'. Importantly, as will become clear in Section 5, the definitions of depth and breadth that we propose in the current paper are quite different from the definitions of depth and width introduced by Mohapatra et al.

## 3. Data

The empirical analyses presented in this paper were carried out using data extracted from the in-house version of the Web of Science (WoS) database available at the Centre for Science and Technology Studies (CWTS) at Leiden University. We made use of the Science Citation Index Expanded, the Social Sciences Citation Index, and the Arts & Humanities Citation Index. We considered only publications of the document types *article*, *review*, and *letter*. Our data covers 36.2 million publications that appeared between 1980 and 2017 and 699.3 million citation relations between these publications.

Our analyses focus on highly cited publications in the period 2000–2017, where a highly cited publication is defined as a publication that has received at least 100 citations at the end of 2017.[1] In total, 550,747 highly cited publications in the period 2000–2017 were identified. For these publications, we calculated the citation impact indicators defined in the next sections. In the calculation of the indicators of dependence and independence, we considered only references to publications included in our data (73.1% of all references). References to publications not included in the data, typically publications not indexed in the WoS database, were not taken into account. This is why our analyses focus on publications from the period 2000–2017 and why publications from the period 1980–1999 are not considered. The calculation of our indicators of dependence and independence for publications from the period 1980–1999 would be affected by the fact that many references in these publications point to literature that appeared before 1980 and that is not included in our data.

Using the algorithmic methodology introduced by Waltman and Van Eck (2012), publications in the WoS database in the period 2000–2017 were clustered based on citation relations. 4,047 clusters of publications were obtained. Clusters are non-overlapping. Each publication belongs to only one cluster. The 4,047 clusters were grouped into the following five broad scientific disciplines:[2]

- Biomedical and health sciences (BHS; 291,342 highly cited publications)
- Life and earth sciences (LES; 73,113 highly cited publications)
- Mathematics and computer science (MCS; 10,475 highly cited publications)
- Physical sciences and engineering (PSE; 148,521 highly cited publications)
- Social sciences and humanities (SSH; 27,149 highly cited publications)

## 4. Level of citation impact

As already mentioned, our citation impact framework distinguishes between three dimensions of the citation impact of a publication, namely (1) level, (2) depth and breadth, and (3) dependence and independence. In this section, we discuss the dimension of the level of citation impact. The depth and breadth and the dependence and independence dimensions are discussed in Sections 5 and 6, respectively.

---

[1] The relative indicators that will be introduced in Sections 5 and 6 provide meaningful results only for publications that have received a substantial number of citations. Most of our analyses therefore focus on highly cited publications.

[2] For more details, see https://www.leidenranking.com/information/fields.

**4.1. Conceptualization and operationalization**

The level of citation impact of a publication reflects how much impact the publication has had on other publications. We operationalize this by the number of citations a publication has received, denoted by CP (i.e., number of citing publications). The larger the number of citations a publication has received, the higher the level of citation impact of the publication. The level of citation impact represents the traditional perspective on the citation impact of a publication.

**4.2. Descriptive statistics**

We now report some descriptive statistics for the CP indicator. Statistics are presented for each of the five broad scientific disciplines and for all disciplines together (labeled 'ALL' in the tables and figures in this paper). The statistics are based on the 550,747 highly cited publications discussed in Section 3. For each discipline, Table 1 reports the median value of the CP indicator. Figure 3 shows the underlying distribution. As expected, the distribution of the CP indicator is highly skewed. The horizontal axis in Figure 3 therefore has a logarithmic scale.

Table 1. Median value of the CP indicator for different disciplines.

|    | BHS | LES | MCS | PSE | SSH | ALL |
|----|-----|-----|-----|-----|-----|-----|
| CP | 150 | 144 | 144 | 149 | 149 | 148 |

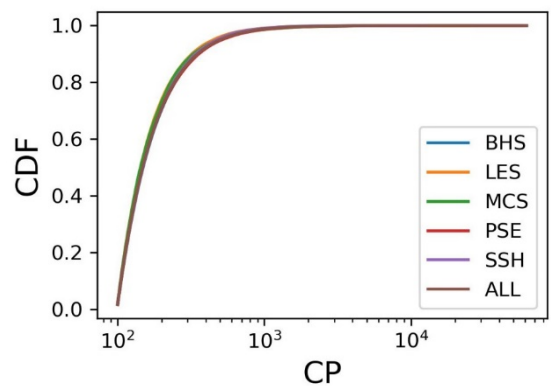

Figure 3. Cumulative distribution function of the CP indicator for different disciplines.

The distribution of the CP indicator is very similar for different disciplines. Normally, to obtain similar citation distributions for different disciplines, a rescaling needs to be performed that normalizes for differences between disciplines in the average level of citation impact (Radicchi et al., 2008; Waltman, Van Eck, & Van Raan, 2012). However, we consider only highly cited publications (i.e., the tail of the citation distribution). For these publications, there turns out to be no need for performing a rescaling.

## 5. Depth and breadth of citation impact

To motivate the idea of the depth and breadth and the dependence and independence of the citation impact of a publication, we consider the following article dealing with a topic in the field of webometrics: Thelwall, M. (2001). Extracting macroscopic information from web links. *Journal of the American Society for Information Science and Technology*, *52*(13), 1157–1168. For simplicity, we refer to this article as publication P. In our data, P cites 43 publications and is cited by 107 publications. Some of the 107 citing publications also cite other publications citing P. For a given publication citing P, R[citing pub] denotes the number of references to other publications citing P. Some publications citing P also cite publications cited by P. For a given publication citing P, R[cited pub] denotes the number of references to publications cited by P.

The plots in Figure 4 show the distributions of R[citing pub] and R[cited pub] for the 107 publications citing P. As can be seen in the left plot, some publications citing P have a high value for R[citing pub]. There even is a publication that cites P and that also cites 42 other publications citing P. However, there are also publications that cite P and that do not cite any other publication citing P. Likewise, the right plot shows that some publications citing P have a high value for R[cited pub]. There is one publication that cites P and that also cites 22 publications cited by P. The other way around, some publications citing P do not cite any publication cited by P.

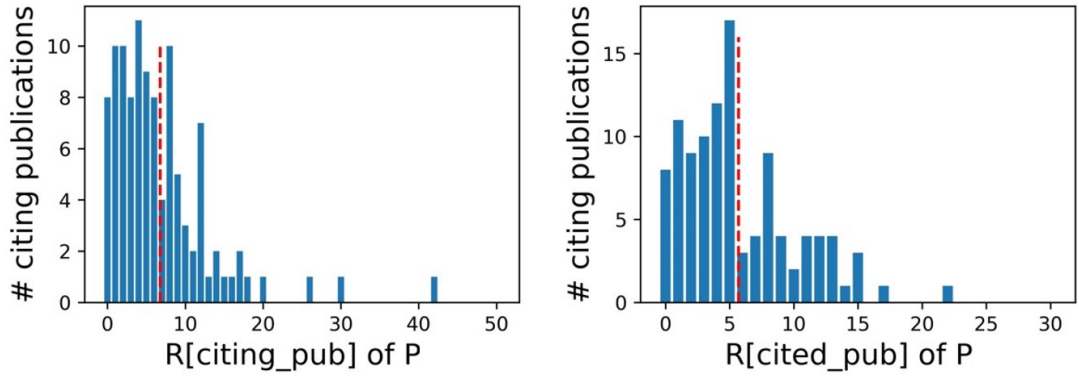

Figure 4. Distributions of R[citing pub] and R[cited pub] for the citing publications of publication P (Thelwall, 2001). Dashed vertical lines indicate the mean of a distribution.

The two distributions discussed in the above example provide important information about the citation impact of a publication. In the rest of this section, we use the distribution of R[citing pub] to quantify the depth and breadth of a publication's citation impact. In Section 6, the distribution of R[cited pub] is used to quantify the dependence and independence of a publication's citation impact.

**5.1. Conceptualization and operationalization**

To understand the notion of the depth and breadth of the citation impact of a publication, we consider an example involving two publications, $A$ and $B$. These publications have received the same number of citations, and they therefore have the same level of citation impact. However, $A$ and $B$ differ in how they have an impact on other publications. Publication $A$ introduces an innovative new idea in a research field. Many publications in this field start to build on this idea. These publications all cite $A$ and many of them also cite each other. In contrast, outside the research field of $A$, little attention is paid to the idea introduced in $A$ and few citations are made to $A$. The situation is very different for $B$. This publication introduces a new software tool for carrying out certain statistical analyses. The tool turns out to be useful in many different research fields. In all these fields, publications that use the tool cite $B$. However, apart from the fact that they use the tool introduced in $B$, these publications have little in common. They all deal with different research problems. In general, publications citing $B$ therefore do not cite each other. In this example, $A$ and $B$ have an impact on other publications in very different ways. We say that $A$ has a deep citation impact while $B$ has a broad citation impact. Figure 1 illustrates the difference between the deep citation impact of *A* and the broad citation impact of *B*.

To quantify the depth and breadth of the citation impact of a publication, we propose the six indicators summarized in Table 2. On the one hand, we distinguish between indicators of depth and indicators of breadth. On the other hand, we also make a distinction between absolute and relative indicators. Absolute indicators scale with the level of citation impact of a publication, while relative indicators are normalized for the level of citation impact. Relative indicators are defined only for publications that have received at least one citation (i.e., $CP > 0$). From a relative point of view, depth and breadth are opposite concepts. A high depth implies a low breadth, and *vice versa*. Hence, when a relative perspective is taken, depth and breadth can be seen as two sides of the same coin. This is different when an absolute perspective is taken. From an absolute point of view, a publication may have both a high depth and a high breadth, or it may have both a low depth and a low breadth. This means that, from an absolute point of view, depth and breadth are conceptually distinct dimensions, even though they may be empirically correlated.

To further illustrate the distinction between absolute and relative perspectives, suppose that we are interested in measuring poverty at the level of countries. Suppose that for each person living in a country we are able to determine whether this person is considered to be poor or not. From an absolute point of view, we can then count the number of poor people and the number of non-poor people in a country. These are two conceptually distinct dimensions. For instance, if there are 100,000 poor people in a country, this does not tell us anything about the number of non-poor people. There may be no non-poor people at all in the country, but there may also be 100 million non-poor people. Now consider the relative point of view. From this point of view, we do not look at the number of poor people and the number of non-poor people in a country, but we look at the proportion of poor people and the proportion of non-poor people. These two proportions are, of course, two sides of the same coin. If we know the proportion of poor people in a country (e.g., 10%), we also know the proportion of non-poor people (e.g., 90%). Hence, the two proportions represent the same conceptual dimensions. Our distinction between absolute and relative indicators of depth and breadth of citation impact is analogous to the distinction between absolute and relative measurements of poverty, but instead of countries and the people living in them, we consider publications and the citations they receive.

Although we propose six indicators of the depth and breadth of the citation impact of a publication, our idea is that practical applications will probably use only one or two of them. We now discuss the six indicators in more detail.

Table 2. Indicators of the depth and breadth of the citation impact of a publication.

|  | Absolute indicators | Relative indicators |
|---|---|---|
| Breadth | **CP(R[citing pub] = 0)**<br>Number of publications citing the focal publication that do not cite other publications citing the focal publication. | **PCP(R[citing pub] = 0)**<br>Proportion of publications citing the focal publication that do not cite other publications citing the focal publication. |
| Depth | **CP(R[citing pub] > 0)**<br>Number of publications citing the focal publication that also cite other publications citing the focal publication. | **PCP(R[citing pub] > 0)**<br>Proportion of publications citing the focal publication that also cite other publications citing the focal publication. |
| Depth | **TR[citing pub]**<br>Total number of references in publications citing the focal publication to other publications citing the focal publication. | **MR[citing pub]**<br>Average number of references in publications citing the focal publication to other publications citing the focal publication. |

CP(R[citing pub] = 0) and PCP(R[citing pub] = 0) denote the number and the proportion of publications citing the focal publication that do not cite other publications citing the focal publication. CP(R[citing pub] = 0) is an indicator of the absolute breadth of the citation impact of the focal publication, while PCP(R[citing pub] = 0) = CP(R[citing pub] = 0)/CP is an indicator of the relative breadth.

Consider publications $A$ and $B$ in Figure 1. CP(R[citing pub] = 0) = 1 for $A$ because $A1$ is the only publication that cites $A$ and that does not cite other publications citing $A$. CP(R[citing pub] = 0) = 5 for $B$ because none of the five publications citing $B$ cites other publications citing $B$. Furthermore, PCP(R[citing pub] = 0) = 1/5 for $A$ and PCP(R[citing pub] = 0) = 5/5 = 1 for $B$. The CP(R[citing pub] = 0) and PCP(R[citing pub] = 0) indicators show that $B$ has a broader citation impact than $A$, both in absolute and in relative terms.

CP(R[citing pub] > 0) and PCP(R[citing pub] > 0) denote the number and the proportion of publications citing the focal publication that also cite other publications citing the focal publication. CP(R[citing pub] > 0) is an indicator of the absolute depth of the citation impact of the focal publication, while PCP(R[citing pub] > 0) = CP(R[citing pub] > 0)/CP is an indicator of the relative depth.

In Figure 1, CP(R[citing pub] > 0) = 4 for $A$ because $A2$, $A3$, $A4$, and $A5$ all cite $A$ and also cite other publications citing $A$. CP(R[citing pub] > 0) = 0 for $B$ because none of the five publications citing $B$ cites other publications citing $B$. Furthermore, PCP(R[citing pub] > 0) = 4/5 for $A$ and PCP(R[citing pub] > 0) = 0/5 = 0 for $B$. The CP(R[citing pub] > 0) and PCP(R[citing pub] > 0) indicators show that $A$ has a deeper citation impact than $B$.

Like CP(R[citing pub] > 0) and PCP(R[citing pub] > 0), TR[citing pub] and MR[citing pub] are indicators of, respectively, the absolute and the relative depth of the citation impact of a publication. TR[citing pub] denotes the total number of references in publications citing the focal publication to other publications citing the focal publication. MR[citing pub] = TR[citing pub]/CP denotes the average number of references in publications citing the focal publication to other publications citing the focal publication.

In Figure 1, TR[citing pub] = 10 for $A$ because there are citation relations between all $(5 \times 4)/2 = 10$ pairs of publications citing $A$. TR[citing pub] = 0 for $B$ because the five publications citing $B$ do not cite each other. Furthermore, MR[citing pub] = 10/5 = 2 for $A$ and MR[citing pub] = 0/5 = 0 for $B$. Like the CP(R[citing pub] > 0) and PCP(R[citing pub] > 0) indicators, the TR[citing pub] and MR[citing pub] indicators show that $A$ has a deeper citation impact than $B$.

Our absolute indicators of depth and breadth are related to indicators proposed by Huang et al. (2018). CP(R[citing pub] = 0) is essentially equivalent to the number of 'isolate endorsers' in the terminology of Huang et al. Likewise, CP(R[citing pub] > 0) is essentially equivalent to the sum of the number of 'late endorsers' and the number of 'connectors'. TR[citing pub] is equivalent to the number of direct citations between citing publications in the terminology of Huang et al. Furthermore, CP(R[citing pub] = 0) is also similar, but not identical, to the citation counts studied by Clough et al. (2015). These citation counts are obtained from the transitive reduction of a citation network.

We do not intend to make a normative judgment by quantifying the depth and breadth of the citation impact of a publication. From our point of view, a deeper citation impact is not necessarily better than a broader citation impact, or the other way around. However, we do believe that the distinction between deep and broad citation impact is useful to get a more detailed understanding of the way in which a publication has an impact on other publications. We will illustrate this in the case study presented in Section 7.

**5.2. Descriptive statistics**

We now report some basic descriptive statistics for our indicators of the depth and breadth of the citation impact of a publication. For each of our broad scientific disciplines, Table 3 reports the median values of both the absolute and the relative indicators. Figure 5 shows the underlying distributions. Because of the skewness of the distributions, the horizontal axes in Figure 5 have a logarithmic scale. Table A1 in the appendix reports correlations between the various indicators.

Based on the indicators of the absolute depth and breadth of citation impact (i.e., $CP(R[\text{citing pub}] = 0)$, $CP(R[\text{citing pub}] > 0)$, and $TR[\text{citing pub}]$), we observe that PSE publications tend to have a relatively deep citation impact, while MCS and SSH publications tend to have a relatively broad citation impact. This seems to suggest that PSE research is of a stronger cumulative nature than MCS and SSH research. The relative indicators yield a similar picture (as is to be expected, since the distribution of the level of citation impact is almost the same for all disciplines; see Subsection 4.2).

Table 3. Median values for different disciplines of the indicators of the depth and breadth of the citation impact of a publication.

| Absolute indicators | | | | | | |
|---|---|---|---|---|---|---|
| | BHS | LES | MCS | PSE | SSH | ALL |
| $CP(R[\text{citing pub}] = 0)$ | 48 | 42 | 53 | 40 | 57 | 46 |
| $CP(R[\text{citing pub}] > 0)$ | 102 | 103 | 93 | 109 | 95 | 104 |
| $TR[\text{citing pub}]$ | 361 | 388 | 277 | 442 | 280 | 378 |
| Relative indicators | | | | | | |
| | BHS | LES | MCS | PSE | SSH | ALL |
| $PCP(R[\text{citing pub}] = 0)$ | 0.30 | 0.27 | 0.34 | 0.24 | 0.35 | 0.28 |
| $PCP(R[\text{citing pub}] > 0)$ | 0.70 | 0.73 | 0.66 | 0.76 | 0.65 | 0.72 |
| $MR[\text{citing pub}]$ | 2.31 | 2.56 | 1.78 | 2.81 | 1.79 | 2.42 |

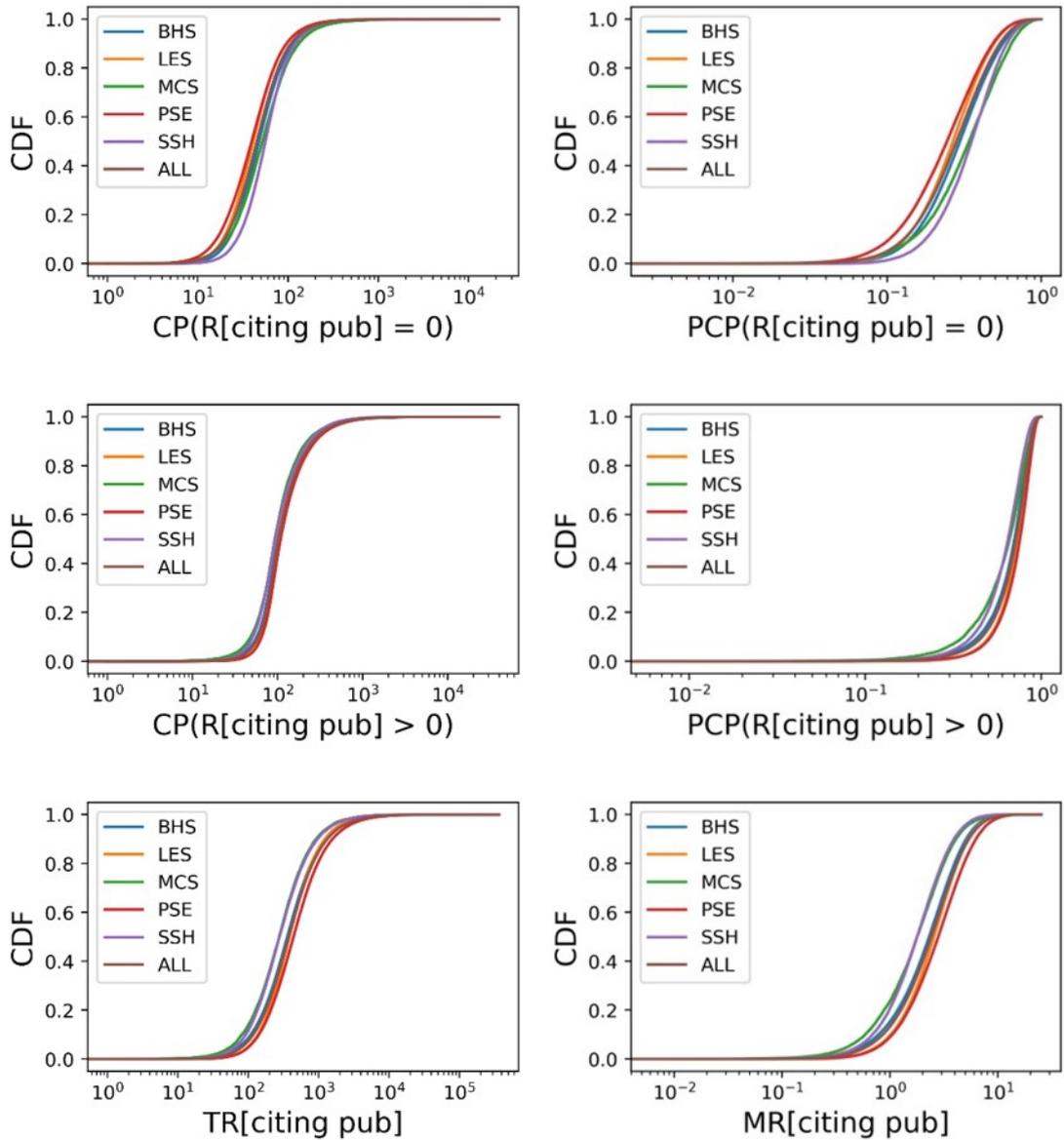

Figure 5. Cumulative distribution functions for different disciplines of the indicators of the depth and breadth of the citation impact of a publication (left: absolute indicators; right: relative indicators).

## 6. Dependence and independence of citation impact

In this section, we consider the dependence and independence of the citation impact of a publication.

### 6.1. Conceptualization and operationalization

Two publications may have a similar level and a similar depth and breadth of citation impact, but nevertheless there may be an important difference in how they have an impact on other publications. Some publications may have an impact by building on earlier publications and by contributing new scientific knowledge in a cumulative way. It is likely that these publications will be cited together with publications that they build on and that they cite. These publications have a citation impact that depends on earlier publications. We therefore say that these publications have a dependent citation impact. This is illustrated by publication *A* in Figure 2. Other publications may have an impact without relying strongly on earlier publications. These publications may introduce new ideas that have been developed relatively independently from earlier literature. These publications usually will not be cited together with publications that they cite. We say that these publications have an independent citation impact. An illustration is provided by publication *B* in Figure 2.

Our operationalization of the dependence and independence of the citation impact of a publication mirrors the operationalization of depth and breadth discussed in Subsection 5.1. Table 4 summarizes the six indicators that we propose for quantifying dependence and independence. Like in the case of depth and breadth, we distinguish between absolute and relative indicators. From a relative point of view, dependence and independence are opposite concepts. A high dependence implies a low independence, and *vice versa*. From an absolute point of view, dependence and independence are conceptually distinct dimensions. From this viewpoint, a publication may for instance have both a high dependence and a high independence.

In practical applications, there usually will be no need to use all six indicators of dependence and independence. A typical application will probably use one or two of them. We now discuss the six indicators in more detail.

Table 4. Indicators of the dependence and independence of the citation impact of a publication.

|  | Absolute indicators | Relative indicators |
| --- | --- | --- |
| Independence | **CP(R[cited pub] = 0)** Number of publications citing the focal publication that do not cite publications cited by the focal publication. | **PCP(R[cited pub] = 0)** Proportion of publications citing the focal publication that do not cite publications cited by the focal publication. |
| Dependence | **CP(R[cited pub] > 0)** | **PCP(R[cited pub] > 0)** |

|  | Number of publications citing the focal publication that also cite publications cited by the focal publication. | Proportion of publications citing the focal publication that also cite publications cited by the focal publication. |
|---|---|---|
| Dependence | **TR[cited pub]** Total number of references in publications citing the focal publication to publications cited by the focal publication. | **MR[cited pub]** Average number of references in publications citing the focal publication to publications cited by the focal publication. |

CP(R[cited pub] = 0) and PCP(R[cited pub] = 0) denote the number and the proportion of publications citing the focal publication that do not cite publications cited by the focal publication. CP(R[cited pub] = 0) is an indicator of the absolute independence of the citation impact of the focal publication, while PCP(R[cited pub] = 0) = CP(R[cited pub] = 0)/CP is an indicator of the relative independence.

Consider publications $A$ and $B$ in Figure 2. CP(R[cited pub] = 0) = 0 for $A$ because all five publications citing $A$ also cite publications cited by $A$. CP(R[cited pub] = 0) = 5 for $B$ because none of the five publications citing $B$ also cites publications cited by $B$. Furthermore, PCP(R[cited pub] = 0) = 0/5 = 0 for $A$ and PCP(R[cited pub] = 0) = 5/5 = 1 for $B$. The CP(R[cited pub] = 0) and PCP(R[cited pub] = 0) indicators show that $B$ has a more independent citation impact than $A$, both in absolute and in relative terms.

CP(R[cited pub] > 0) and PCP(R[cited pub] > 0) denote the number and the proportion of publications citing the focal publication that also cite publications cited by the focal publication. CP(R[cited pub] > 0) is an indicator of the absolute dependence of the citation impact of the focal publication, while PCP(R[cited pub] > 0) = CP(R[cited pub] > 0)/CP is an indicator of the relative dependence.

In Figure 2, CP(R[cited pub] > 0) = 5 for $A$ because all five publications citing $A$ also cite publications cited by $A$. CP(R[cited pub] > 0) = 0 for $B$ because none of the five publications citing $B$ also cites publications cited by $B$. Furthermore, PCP(R[cited pub] > 0) = 5/5 = 1 for $A$ and PCP(R[citing pub] > 0) = 0/5 = 0 for $B$. The CP(R[cited pub] > 0) and PCP(R[cited pub] > 0) indicators show that $A$ has a more dependent citation impact than $B$.

Like CP(R[cited pub] > 0) and PCP(R[cited pub] > 0), TR[cited pub] and MR[cited pub] are indicators of, respectively, the absolute and the relative dependence of the citation impact of a publication. TR[cited pub] denotes the total number of references in publications citing the focal publication to publications cited by the focal publication. MR[cited pub] = TR[cited pub]/CP denotes the average number of references in publications citing the focal publication to publications cited by the focal publication.

In Figure 2, TR[cited pub] = 15 for $A$ because there are citation relations between all $5 \times 3 = 15$ pairs of a publication citing $A$ and a publication cited by $A$. TR[cited pub] = 0 for $B$ because the five publications citing $B$ do not cite publications cited by $B$. Furthermore, MR[cited pub] = 15/5 = 3 for $A$ and MR[cited pub] = 0/5 = 0 for $B$. Like the CP(R[cited pub] > 0) and PCP(R[cited pub] > 0) indicators, the TR[cited pub] and MR[cited pub] indicators show that $A$ has a more dependent citation impact than $B$.

Our absolute indicators of dependence and independence are related to statistics studied by Wu et al. (2019). CP(R[cited pub] > 0) and CP(R[cited pub] = 0) are equivalent to, respectively, $n_j$ and $n_i$ in Figure 1 in Wu et al.

**6.2. Descriptive statistics**

We now report some basic descriptive statistics for our indicators of the dependence and independence of the citation impact of a publication. For each of our broad scientific disciplines, Table 5 reports the median values of both the absolute and the relative indicators. Figure 6 shows the underlying distributions. Like in Figures 4 and 5, the horizontal axes have a logarithmic scale. Table A2 in the appendix reports correlations between the various indicators.

As can be seen in Table 5 and Figure 6, MCS publications have a relatively independent citation impact, both from an absolute and from a relative viewpoint. Compared with publications in other disciplines, MCS publications have a citation impact that is less dependent on earlier publications. However, this may partly be an artifact of our data. As explained in Section 3, references to publications not included in our data are disregarded. Publications in conference proceedings, which play an important role in MCS, are not included in our data. This may artificially decrease the dependence of the citation impact of MCS publications.

Table 5. Median values for different disciplines of the indicators of the dependence and independence of the citation impact of a publication.

| Absolute indicators |
|---|

|                      | BHS  | LES  | MCS  | PSE  | SSH  | ALL  |
|----------------------|------|------|------|------|------|------|
| CP(R[cited pub] = 0) | 44   | 40   | 62   | 41   | 46   | 43   |
| CP(R[cited pub] > 0) | 107  | 105  | 88   | 109  | 104  | 107  |
| TR[cited pub]        | 402  | 375  | 191  | 399  | 328  | 390  |
| Relative indicators  |      |      |      |      |      |      |
|                      | BHS  | LES  | MCS  | PSE  | SSH  | ALL  |
| PCP(R[cited pub] = 0)| 0.27 | 0.26 | 0.40 | 0.25 | 0.28 | 0.27 |
| PCP(R[cited pub] > 0)| 0.73 | 0.74 | 0.60 | 0.75 | 0.72 | 0.73 |
| MR[cited pub]        | 2.43 | 2.34 | 1.20 | 2.39 | 1.99 | 2.36 |

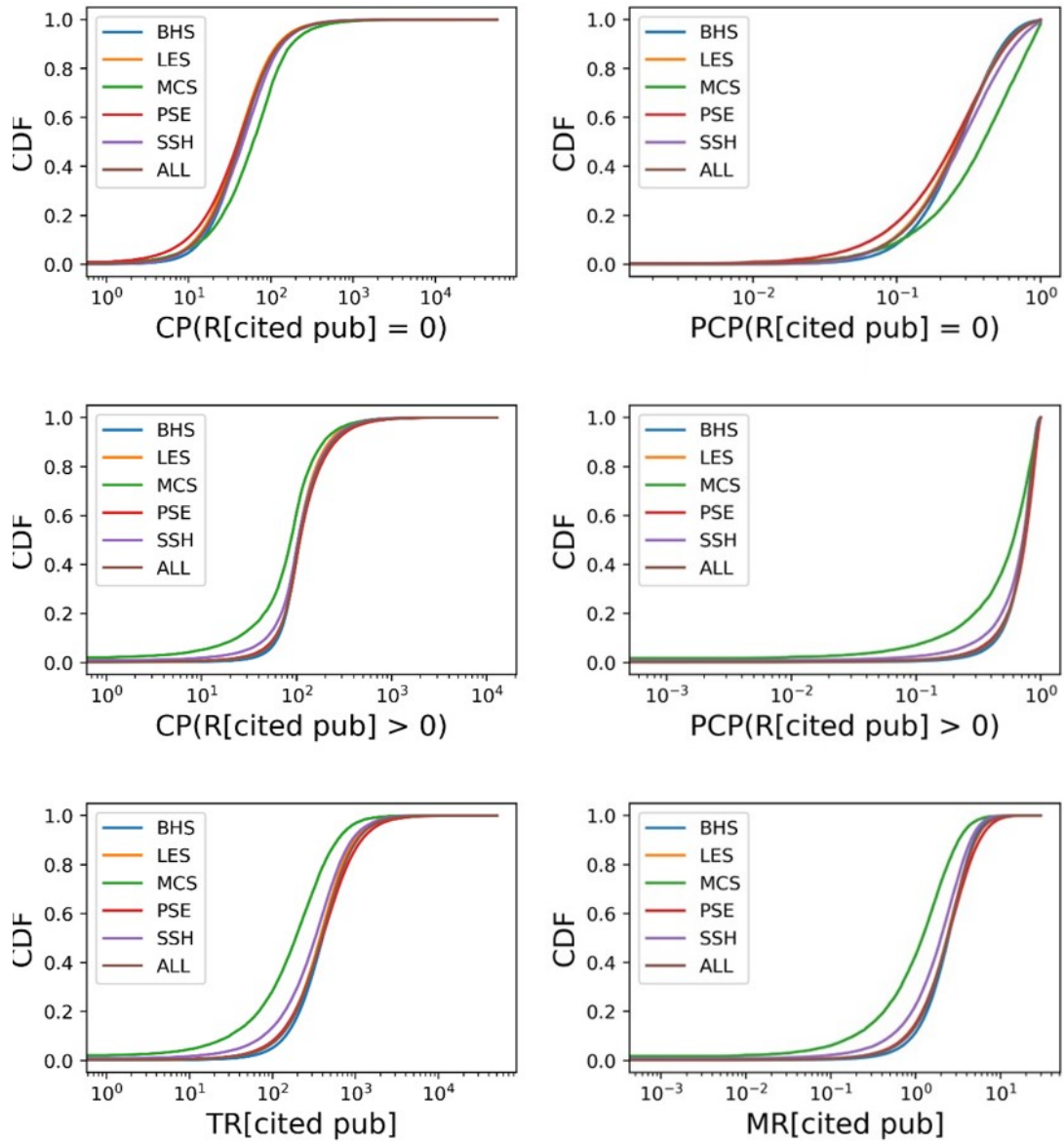

Figure 6. Cumulative distribution functions for different disciplines of the indicators of the dependence and independence of the citation impact of a publication (left: absolute indicators; right: relative indicators).

## 7. Case study in the field of scientometrics

To demonstrate the value of our multi-dimensional framework for characterizing the citation impact of publications, we now present a case study in which the framework is applied to publications in the field of scientometrics. As explained in Section 3, using an algorithmic methodology, 4,047 clusters of publications were obtained, covering all scientific disciplines. One of these clusters can be considered to represent the field of scientometrics. We selected the 14,464 publications in this cluster. This includes 182 highly cited publications, each of which has received at least 100 citations. We calculated our citation impact indicators for all 14,464 publications. Below, we first discuss the absolute indicators and then the relative ones.

**7.1. Absolute indicators**

Figure 7 presents scatter plots showing the correlation between the level of citation impact of the scientometrics publications and the absolute depth and breadth and the absolute dependence and independence of the citation impact of these publications. We use $CP(R[\text{citing pub}] > 0)$ and $CP(R[\text{citing pub}] = 0)$ as indicators of, respectively, absolute depth and absolute breadth, and $CP(R[\text{cited pub}] > 0)$ and $CP(R[\text{cited pub}] = 0)$ as indicators of absolute dependence and absolute independence. For each of the indicators, Table A3 in the appendix lists the top 10 publications. Based on Figure 7 and Table A3, we observe a substantial correlation between the indicators. This is to be expected, since absolute indicators all depend on the number of citations a publication has received. Nevertheless, there are also important differences between the indicators.

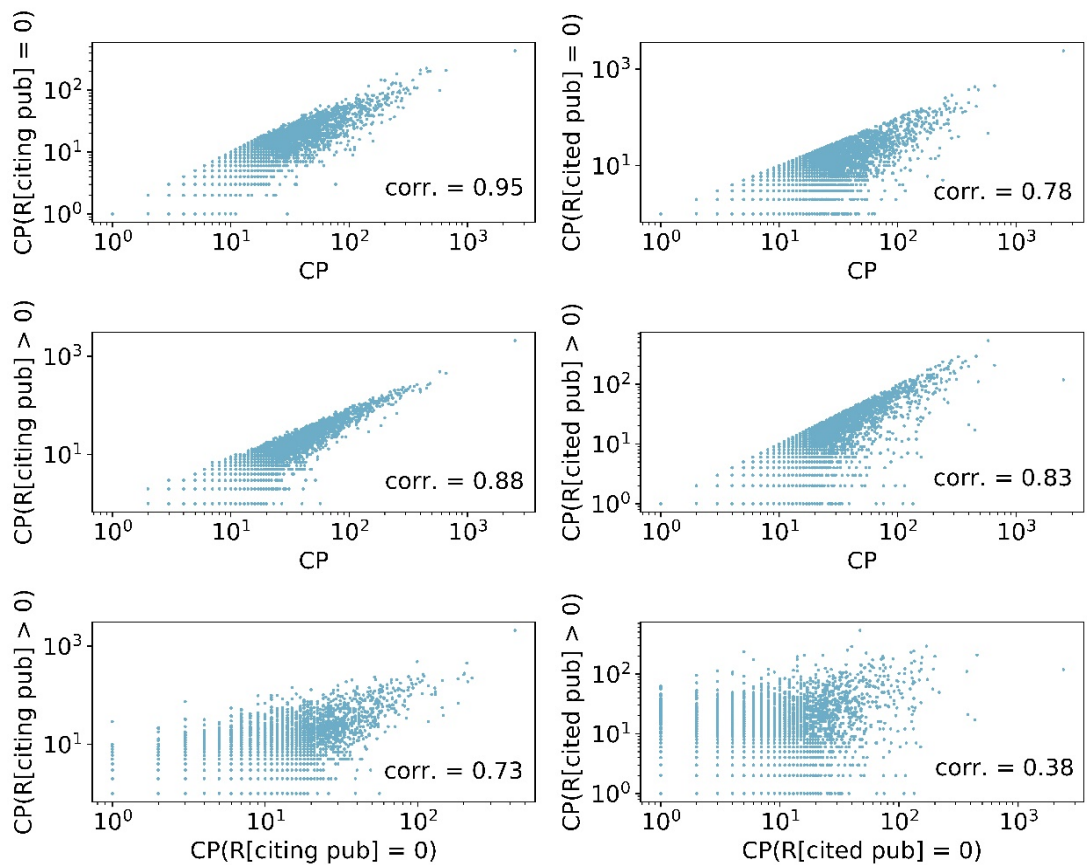

Figure 7. Correlation between the level, the absolute depth and breadth (left), and the absolute dependence and independence (right) of the citation impact of scientometrics publications. Spearman's rank correlation coefficient is used to quantify the strength of the correlation.

The article by Egghe (2006), in which the so-called g-index was introduced as an alternative to the well-known h-index, offers a clear illustration of these differences. As can be seen in Table A3, the CP indicator shows that this is the third most cited publication in the field of scientometrics. In terms of the absolute depth of citation impact, this publication even ranks second, while it ranks first in terms of the absolute dependence of citation impact. The prominent ranking of Egghe's article in terms of absolute depth and absolute dependence can be explained by the important contribution made by this publication to a large stream of publications dealing with the h-index and other bibliometric indicators of the performance of individual researchers. Many of these publications cite each other, while they also often cite Egghe's article as well as the article by Hirsch (2005) in which the h-index was proposed. Looking at the top 10 publications based on absolute breadth and absolute independence, it turns out the Egghe's article is not even included. Hence, Egghe's article has a deep citation impact, but its citation impact is not very broad. Also, the citation impact of Egghe's article is strongly dependent on the citation impact of the article by Hirsch. The independent citation impact of Egghe's article is therefore limited.

Another example of a publication for which the different indicators yield quite different results is the article by Falagas et al. (2008), in which the strengths and weaknesses of a number of bibliographic databases are discussed. As shown in Table A3, this article is ranked tenth in terms of absolute depth, while it is ranked second in terms of absolute breadth. Hence, contrary to Egghe's article discussed above, the article by Falagas et al. has a very broad but not so deep citation impact. In other words, the article has been cited a lot, but many of the citing publications do not seem to be part of a coherent body of literature. Presumably, many researchers cite the article to explain why they use a specific bibliographic database, without engaging more substantively with the article. The lack of more substantive engagement may be partly due to the fact that the article was published in a life sciences journal, not in a scientometric journal. It is also remarkable that the article by Falagas et al. is ranked third in terms of absolute independence, while it is not included at all in the top 10 publications based on absolute dependence. Again, this is opposite to what we observed for the article by Egghe. The independence of the citation impact of the article by Falagas et al. seems to reflect the fact that this was one of the first publications in which bibliographic databases were compared.

As already mentioned, the different absolute indicators are quite strongly correlated with each other, since they all depend on the number of citations a publication has received. We now turn to relative indicators, for which the differences can be expected to be more substantial.

**7.2. Relative indicators**

Relative indicators provide meaningful results only for publications that have received a substantial number of citations. In our analysis based on relative indicators, we therefore consider only the 182 scientometrics publications that have received at least 100 citations.[3] Figure 8 presents two scatter plots that both show the relative depth and the relative dependence of the citation impact of the 182 highly cited publications.

---

[3] The full results of the analysis are available in a data repository (Bu & Waltman, 2020).

In the top plot, PCP(R[citing pub] > 0) and PCP(R[cited pub] > 0) are used as indicators of, respectively, relative depth and relative dependence. In the bottom plot, we instead use the MR[citing pub] and MR[cited pub] indicators. We do not consider indicators of relative breadth and relative independence. From a relative point of view, breadth is the direct opposite of depth and likewise independence is the direct opposite of dependence.[4] Therefore it is sufficient to look only at depth and dependence.

---

[4] We emphasize that this is the case only when a relative point of view is taken. As discussed in Subsections 5.1 and 6.1, this is not the case when an absolute point of view is taken.

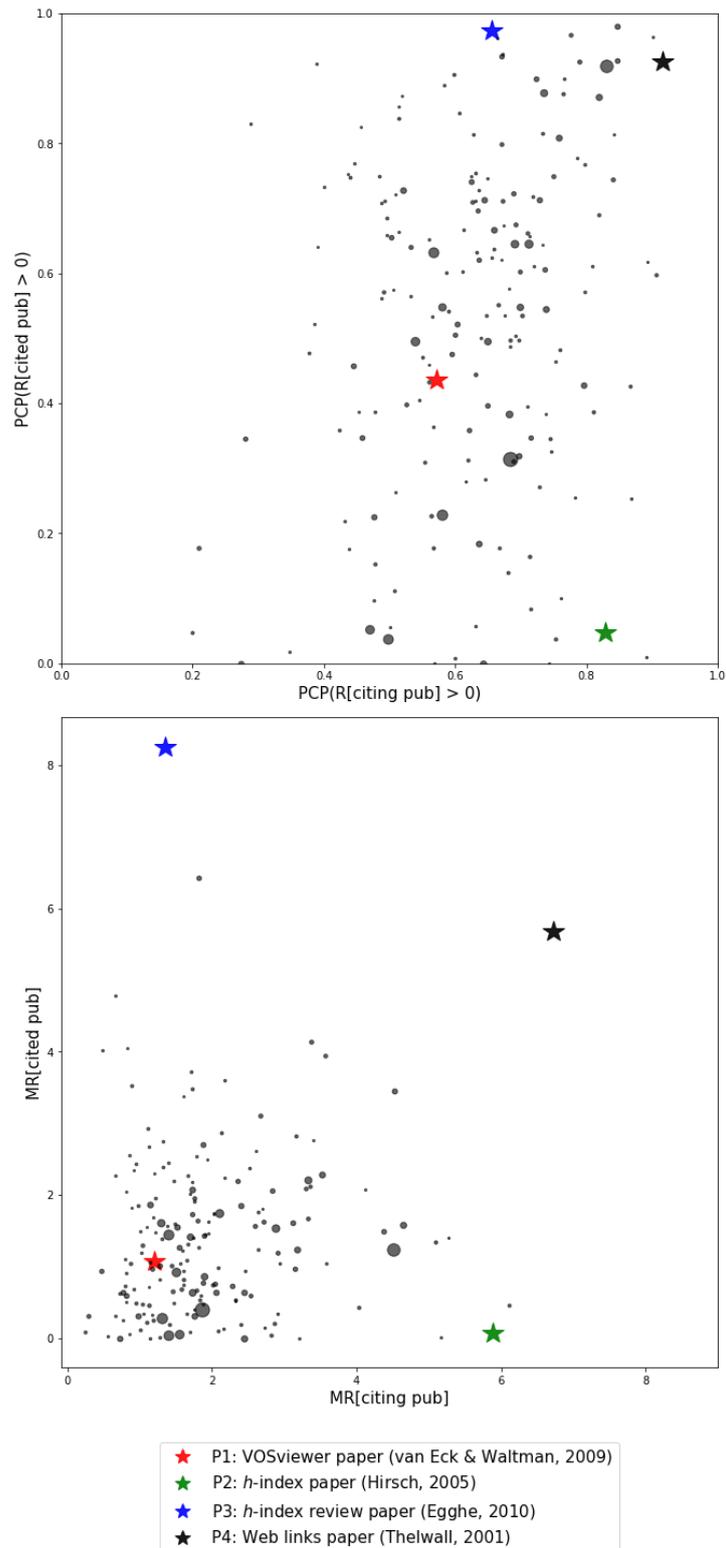

Figure 8. Correlation between the relative depth (horizontal axis) and the relative dependence (vertical axis) of the citation impact of highly cited scientometrics publications. The two plots show different indicators of depth and dependence. P1, P2, P3, and P4 denote four publications selected for a more detailed analysis.

The four publications denoted by P1, P2, P3, and P4 in Figure 8 were selected for a more detailed analysis. We selected a publication with a low depth and a low dependence (P1), a publication with a high depth and a low dependence (P2), a publication with a low depth and a high dependence (P3), and a publication with a high depth and a high dependence (P4). We chose publications with which we are sufficiently familiar ourselves, so that we are able to offer a detailed interpretation of the citation impact of the selected publications.

Table 6 lists the four selected publications and reports their number of citations and number of references. In addition, for each of the selected publications, Figure 9 shows the distribution of the number of citations from publications citing the selected publication to other publications citing the selected publication (i.e., the distribution of R[citing pub]) as well as the distribution of the number of citations from publications citing the selected publication to publications cited by the selected publication (i.e., the distribution of R[cited pub]). We note that publication P4 is identical to publication P discussed in Section 5. Using the information provided in Figures 8 and 9 and Table 6, we now offer an in-depth interpretation of the citation impact of the four selected publications.

Table 6. The four selected publications and their number of citations and number of references.

|  | P1 | P2 | P3 | P4 |
|---|---|---|---|---|
| Authors | N. J. van Eck & L. Waltman | J. E. Hirsch | L. Egghe | M. Thelwall |
| Title | Software survey: VOSviewer, a computer program for bibliometric mapping | An index to quantify an individual's scientific research output | The Hirsch index and related impact measures | Extracting macroscopic information from Web links |
| Journal | Scientometrics | PNAS | Ann. Rev. of Inf. Sc. and Tech. | J. of Am. Soc. for Inf. Sc. and Tech. |
| Publication year | 2009 | 2005 | 2010 | 2001 |
| # cit. | 273 | 2518 | 116 | 107 |
| # ref. | 37 | 6 | 256 | 65 |
| # ref. in data | 26 | 4 | 175 | 43 |

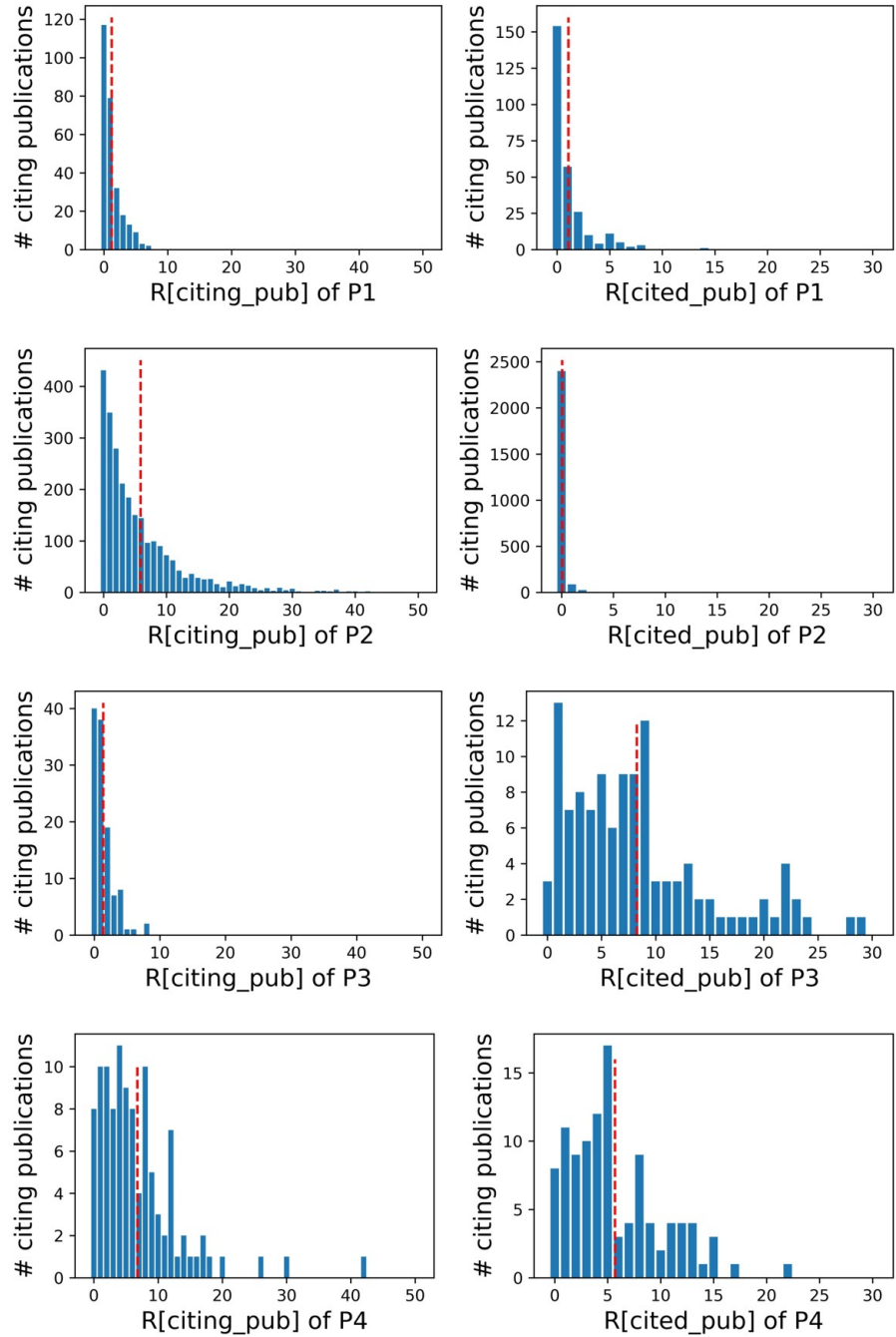

Figure 9. Distributions of R[citing pub] and R[cited pub] for the citing publications of the four selected publications. Dashed vertical lines indicate the mean of a distribution.

Publication P1 is the article that introduced the popular VOSviewer software for visualizing bibliometric networks (Van Eck & Waltman, 2009). VOSviewer is used in a large number of publications in many different research fields. Publications that use VOSviewer often cite P1. In our data, P1 has been cited 273 times. Publications that use VOSviewer typically present a bibliometric analysis of the scientific literature in a specific research field or on a specific research topic. Such publications use VOSviewer as a tool for bibliometric visualization. They usually do not aim to develop new bibliometric methods or tools. Consequently, most publications citing P1 do not contribute to the methodological literature on bibliometric visualization. Publications citing P1 therefore tend to refer only sparsely to other publications on bibliometric visualization. This is reflected by the relatively low depth and dependence of P1. MR[citing pub] equals 1.19, which indicates that on average a publication citing P1 also cites 1.19 other publications citing P1. This means that publications citing P1 are only weakly connected by citation relations. It shows that P1 does not have a very deep citation impact. MR[cited pub] equals 1.07. Hence, when a publication cites P1, on average it also cites 1.07 publications cited by P1. Figure 9 shows that there are a few publications citing P1 that have somewhat more substantial values for R[citing pub] or R[cited pub]. Unlike most publications citing P1, these may be publications that contribute to the methodological literature on bibliometric visualization.

Publication P2 is the article by Hirsch (2005) in which he introduced the h-index. This is an extremely influential publication. With 2518 citations, P2 is by far the most highly cited scientometrics publication in our data. There are a large number of publications that present studies of the h-index, propose alternatives to the h-index, or report bibliometric analyses in which the h-index is applied. In the field of scientometrics, P2 arguably can be seen as the starting point of a new subfield of research focused on studying bibliometric indicators of the performance of individual researchers (or, alternatively, one may suggest there has been an h-bubble; see Rousseau, García-Zorita, & Sanz-Casado, 2013). MR[citing pub] equals 5.89 for P2. Hence, on average, publications that cite P2 also cite 5.89 other publications citing P2. This shows that publications citing P2 are strongly connected by citation relations, which reflects the central position of P2 in a highly active subfield of research. The dependence of P2 is very low. MR[cited pub] equals 0.06, indicating that publications citing P2 hardly cite any publications cited by P2. This suggests that P2 does not only have a central position in a specific subfield of research, but that it can be considered a foundational publication in this subfield. However, P2 has only a very limited number of references (see Table 6), which means that it has a low dependence almost by necessity. The small number of references of P2 can be seen as additional evidence of the foundational role of this publication, but alternatively it may also be argued to reflect a lack of generosity in the referencing behavior of the author of P2.

Publication P3 is a review article about the h-index and other related bibliometric indices (Egghe, 2010). P3 has been cited 116 times in our data. It has 256 references, of which 175 point to publications included in our data (see Table 6). The large number of references reflects the voluminous literature on the h-index published between 2005 and 2010. P3 has a high dependence. MR[cited pub] equals 8.26. Hence, when a publication cites P3, on average it also cites 8.26 publications cited by P3. As can be seen in Figure 9, some publications citing P3 even cite more than 20 publications cited by P3. The high dependence of P3 indicates that P3 builds on a large body of literature and that the citation impact of P3 is strongly dependent on this literature. This reflects that, as a review article, P3 does not make an original scientific contribution. It is sometimes suggested that researchers tend to cite review articles instead of citing the underlying original works, but the high dependence of P3 shows that this is not the case for P3. MR[citing pub] equals 1.34. On average, a publication that cites P3 also cites 1.34 other publications citing P3, indicating that publications citing P3 are only relatively weakly connected by citation relations. This may be due to the gradual decline in the interest of the scientometric community in the h-index. It also shows that P3 has not developed into a canonical reference for publications dealing with the h-index. This may partly be explained by the fact that around 2010 a number of review articles about the h-index were published more or less at the same time.

Publication P4 is about the extraction of macroscopic information from Web links (Thelwall, 2001). This publication deals with a topic in the field of webometrics, which partly overlaps with the field of scientometrics. P4 was published in 2001. It has received 107 citations in our data. As can be seen in Figure 8, P4 is a quite unique publication in the scientometric literature, because it combines a high depth (i.e., MR[citing pub] = 6.75) with a high dependence (i.e., MR[cited pub] = 5.68). This means that publications citing P4 have lots of citation relations both with each other and with publications cited by P4. As can be seen in Table 6, the number of references of P4 is not exceptionally large, making the high dependence of P4 even more noteworthy. The high depth of P4 suggests that P4 makes an important contribution to a relatively narrow but densely connected area of research. On the other hand, the high dependence of P4 seems to indicate that P4 should not be regarded as a pioneering publication. The citation impact of P4 is strongly dependent on earlier publications. Hence, P4 can be considered to make an important incremental contribution, but not a highly innovative one.

## 8. Discussion and conclusion

### 8.1. Summary

We have proposed a multi-dimensional framework for characterizing the citation impact of scientific publications. Our framework makes a distinction between (1) the level, (2) the depth and breadth, and (3) the dependence and independence of the citation impact of a publication. The level of citation impact is quantified by the number of citations a publication has received. The depth and breadth of the citation impact of a publication are operationalized based on citations from publications citing the focal publication to other publications citing the focal publication. The other way around, the dependence and independence of the citation impact of a publication are operationalized based on citations from publications citing the focal publication to publications cited by the focal publication. Our proposed framework also distinguishes between an absolute and a relative perspective on the depth and breadth and the dependence and independence of citation impact. The absolute perspective scales with the level of citation impact, while the relative perspective normalizes for the level of citation impact.

In a traditional one-dimensional perspective on citation impact, the number of citations received by a publication is used as an indicator of the impact of the publication on later publications. Our multi-dimensional framework offers a more detailed understanding of the citation impact of a publication. It makes a distinction between publications that have a deep citation impact, typically in a relatively narrow research area, and publications that have a broad citation impact, probably covering a wider area of research. It also distinguishes between publications that are strongly dependent on earlier work and publications that make a more independent scientific contribution.

In a case study focusing on the field of scientometrics, we have demonstrated the value of our proposed framework for characterizing the citation impact of publications. From a relative point of view (i.e., after normalizing for the level of citation impact), we found that the article in which the h-index was introduced has a high depth and a low dependence. This reflects the role of this article as the starting point of a new subfield of research within the field of scientometrics. On the other hand, a review article on the h-index has a high dependence, which shows the strong reliance of this article on earlier works. A high dependence can be expected to be a typical feature of review articles. The article in which the VOSviewer software was introduced has a low depth, reflecting that it has a broad rather than a deep citation impact. Finally, an article in the field of webometrics has a high depth and a high dependence, indicating that this article contributes to a strongly cumulative research area, but that it does not play a pioneering role in this area.

**8.2. Applications**

The ideas introduced in this paper may have all kinds of applications, for instance in research assessment and scientific literature search.

In the context of research assessment, our proposed multi-dimensional citation impact framework offers new information for evaluating publications. Two publications that have received a similar number of citations may have quite different characteristics in terms of the depth and breadth and the dependence and independence of their citation impact. This information may be provided to research evaluators in a process of peer review. For instance, when evaluators assess the publication output of a researcher, the information may help draw the evaluators' attention to publications with special characteristics, such as a very broad or a very deep citation impact. Evaluators may then choose to study these publications in more detail.

In the context of scientific literature search, our proposed framework may facilitate alternative ways of presenting search results. In many literature search systems, publications can be ranked based on their number of citations. Indicators of absolute depth and breadth and absolute dependence and independence may be used as alternative criteria for ranking publications. Another possibility is to use these indicators to assign badges to publications, for instance to highlight publications that have an exceptionally broad or an exceptionally deep citation impact. In this way, these publications can be given special emphasis in the presentation of search results.

We acknowledge that the above applications of the ideas presented in this paper require additional research. For specific applications, some of our proposed indicators may turn out to be more useful than others. The indicators may also require additional fine-tuning to optimize them for a specific use case.

**8.3. Future research**

There are many directions for future research. First of all, additional case studies can be carried out to assess the usefulness and validity of our proposed multi-dimensional citation impact framework. Such case studies could also analyze how the proposed indicators change over time for individual publications, and how such changes relate to the accumulation of citations for a given focal publication. Also, the proposed framework can be extended in various ways, for instance by taking into account publication type (e.g., review articles), citation type (e.g., self-citations), and citation context (e.g., location in the full text of the citing publication). For instance, if we omit self-citations, how does this affect the values of the indicators? And how do the values of the indicators differ between review articles and regular articles?

Ideas similar to the ones proposed in this paper can also be explored at aggregate levels rather than at the level of individual publications. For instance, based on the current framework, indicators of depth can be defined at the level of authors instead of publications. One approach could be to first determine the depth of each publication of an author and to then aggregate the outcomes from the publication level to the author level. Another approach could be to consider an author-author citation network and to determine the depth of an author based on this network.

Finally, the distinction between cumulative research and more independent research can be studied in alternative ways. Research areas that are of a strongly cumulative nature for instance may be identified by searching for densely connected subnetworks in a citation network.

**Acknowledgments**

We are grateful to Lutz Bornmann and to anonymous reviewers for their helpful comments on our work. Yi Bu also would like to thank Ying Ding, Yong-Yeol Ahn, Johan Bollen, Staša Milojević, Cassidy R. Sugimoto, Dashun Wang, Xianlei Dong, and Jian Xu for their feedback. An earlier version of this paper was presented at the 17[th] International Conference on Scientometrics and Informetrics (ISSI 2019).

## Author contributions

## Competing interests

## Funding information

Yi Bu acknowledges financial support from the National Natural Science Foundation of China (No. 71904081) and the Chinese Education Department Research Foundation for Humanities and Social Sciences (No. 19YJC870017)

## Data availability

The data used in this paper was obtained from the Web of Science database. We are not allowed to redistribute the data. However, a subset of the data used in Section 7 is available in Zenodo (Bu & Waltman, 2020).

## References

Bornmann, L., Devarakonda, S., Tekles, A., & Chacko, G. (2020a). Are disruption index indicators convergently valid? The comparison of several indicator variants with assessments by peers. *Quantitative Science Studies*, DOI: 10.1162/qss_a_00068.

Bornmann, L., Devarakonda, S., Tekles, A., & Chacko, G. (2020b). Disruptive papers published in *Scientometrics*: Meaningful results by using an improved variant of the disruption index originally proposed by Wu, Wang, and Evans (2019). *Scientometrics*, *123*(2), 1149–1155.

Bornmann, L., & Tekles, A. (2019a). Disruption index depends on length of citation window. *El profesional de la información, 28*(2), e280207.

Bornmann, L., & Tekles, A. (2019b). Disruptive papers published in *Scientometrics*. *Scientometrics, 120*(1), 331–336.

Bu, Y., & Waltman, L. (2020). *A multidimensional framework for characterizing the citation impact of scientific publications* [Data set]. Zenodo. https://doi.org/10.5281/zenodo.4279666

Chen, P., Xie, H., Maslov, S., & Redner, S. (2007). Finding scientific gems with Google's PageRank algorithm. *Journal of Informetrics*, *1*(1), 8–15.

Clough, J. R., Gollings, J., Loach, T. V., & Evans, T. S. (2015). Transitive reduction of citation networks. *Journal of Complex Networks, 3*(2), 189–203.

Ding, Y., Liu, X., Guo, C., & Cronin, B. (2013). The distribution of references across texts: Some implications for citation analysis. *Journal of Informetrics, 7*(3), 583–592.

Egghe, L. (2010). The Hirsch index and related impact measures. *Annual Review of Information Science and Technology, 44*(1), 65–114.

Funk, R. J., & Owen-Smith, J. (2017). A dynamic network measure of technological change. *Management Science, 63*(3), 791–817.

Hirsch, J. E. (2005). An index to quantify an individual's scientific research output. *Proceedings of the National Academy of Sciences of the United States of America, 102*(46), 16569–16572.

Huang, Y., Bu, Y., Ding, Y., & Lu, W. (2018). Number versus structure: Towards citing cascades. *Scientometrics*, *117*(3), 2177–2193.

Huang, Y., Bu, Y., Ding, Y., & Lu, W. (2020). Exploring direct citations between citing publications. *Journal of Information Science*, DOI: 10.1177/0165551520917654.

Mohapatra, D., Maiti, A., Bhatia, S., & Chakraborty, T. (2019). Go wide, go deep: Quantifying the impact of scientific papers through influence dispersion trees. In *2019 ACM/IEEE Joint Conference on Digital Libraries (JCDL)* (pp. 305–314). IEEE.

Radicchi, F., Fortunato, S., & Castellano, C. (2008). Universality of citation distributions: Toward an objective measure of scientific impact. *Proceedings of the National Academy of Sciences of the United States of America, 105*(45), 17268–17272.

Rousseau, R., García-Zorita, C., & Sanz-Casado, E. (2013). The h-bubble. *Journal of Informetrics*, *7*(2), 294–300.

Shibayama, S., & Wang, J. (2020). Measuring originality in science. *Scientometrics, 122*(1), 409–427.

Thelwall, M. (2001). Extracting macroscopic information from web links. *Journal of the American Society for Information Science and Technology, 52*(13), 1157–1168.

Van Eck, N. J., & Waltman, L. (2010). Software survey: VOSviewer, a computer program for bibliometric mapping. *Scientometrics, 84*(2), 523–538.

Walker, D., Xie, H., Yan, K.-H., & Maslov, S. (2007). Ranking scientific publications using a model of network traffic. *Journal of Statistical Mechanics: Theory and Experiment*, 6, P06010.

Waltman, L. (2016). A review of the literature on citation impact indicators. *Journal of Informetrics*, *10*(2), 365–391.

Waltman, L., & Van Eck, N. J. (2012). A new methodology for constructing a publication-level classification system of science. *Journal of the American Society for Information Science and Technology*, *63*(12), 2378–2392.

Waltman, L., & Van Eck, N. J. (2019). Field normalization of scientometric indicators. In W. Glänzel et al. (Eds.), *Handbook of science and technology indicators* (pp. 281-300). Springer.

Waltman, L., Van Eck, N. J., Van Leeuwen, T. N., Visser, M. S., & Van Raan, A. F. J. (2011). Towards a new crown indicator: Some theoretical considerations. *Journal of Informetrics*, *5*(1), 37–47.

Waltman, L., Van Eck, N. J., & Van Raan, A. F. J. (2012). Universality of citation distributions revisited. *Journal of the American Society for Information Science and Technology*, *63*(1), 72–77.

Waltman, L., & Yan, E. (2014). PageRank-related methods for analyzing citation networks. In Y. Ding, R. Rousseau, & D. Wolfram (Eds.), *Measuring scholarly impact: Methods and practice* (pp. 83–100). Springer.

Wan, X., & Liu, F. (2014). WL-index: Leveraging citation mention number to quantify an individual's scientific impact. *Journal of the Association for Information Science and Technology, 65*(12), 2509–2517.

Wu, L., Wang, D., & Evans, J. A. (2019). Large teams develop and small teams disrupt science and technology. *Nature, 566*(7744), 378.

Wu, Q., & Yan, Z. (2019). *Solo citations, duet citations, and prelude citations: New measures of the disruption of academic papers*. arXiv:1905.03461.

Zhu, X., Turney, P., Lemire, D., & Vellino, A. (2015). Measuring academic influence: Not all citations are equal. *Journal of the Association for Information Science and Technology, 66*(2), 408–427.

# Appendix

Table A1. Spearman's rank correlation coefficients for indicators of the depth and breadth of the citation impact of publications (all disciplines).

|  | CP(R[citing pub] = 0) | CP(R[citing pub] > 0) | PCP(R[citing pub] = 0) | PCP(R[citing pub] > 0) | TR[citing pub]) | MR[citing pub]) |
|---|---|---|---|---|---|---|
| CP(R[citing pub] = 0) | 1.00 | 0.07 | 0.69 | -0.69 | -0.22 | -0.62 |
| CP(R[citing pub] > 0) | 0.07 | 1.00 | -0.61 | 0.61 | 0.90 | 0.60 |
| PCP(R[citing pub] = 0) | 0.69 | -0.61 | 1.00 | -1.00 | -0.81 | -0.94 |
| PCP(R[citing pub] > 0) | -0.69 | 0.61 | -1.00 | 1.00 | 0.81 | 0.94 |
| TR[citing pub]) | -0.22 | 0.90 | -0.81 | 0.81 | 1.00 | 0.86 |
| MR[citing pub]) | -0.62 | 0.60 | -0.94 | 0.94 | 0.86 | 1.00 |

Table A2. Spearman's rank correlation coefficients for indicators of the dependence and independence of the citation impact of publications (all disciplines).

|  | CP(R[cited pub] = 0) | CP(R[cited pub] > 0) | PCP(R[cited pub] = 0) | PCP(R[cited pub] > 0) | TR[cited pub]) | MR[cited pub]) |
|---|---|---|---|---|---|---|
| CP(R[cited pub] = 0) | 1.00 | 0.04 | 0.84 | -0.84 | -0.26 | -0.73 |
| CP(R[cited pub] > 0) | 0.04 | 1.00 | -0.46 | 0.46 | 0.84 | 0.41 |
| PCP(R[cited pub] = 0) | 0.84 | -0.46 | 1.00 | -1.00 | -0.67 | -0.87 |
| PCP(R[cited pub] > 0) | -0.84 | 0.46 | -1.00 | 1.00 | 0.67 | 0.87 |
| TR[cited pub]) | -0.26 | 0.84 | -0.67 | 0.67 | 1.00 | 0.78 |
| MR[cited pub]) | -0.73 | 0.41 | -0.87 | 0.87 | 0.78 | 1.00 |

Table A3. Top 10 scientometrics publications ranked by five citation impact indicators: CP (i.e., level), CP(R[citing pub] > 0) (i.e., absolute depth), CP(R[citing pub] = 0) (i.e., absolute breadth), CP(R[cited pub] > 0) (i.e., absolute dependence), and CP(R[cited pub] = 0) (i.e., absolute independence).

| Ranked by CP | | | | | | | | |
|---|---|---|---|---|---|---|---|---|
| CP | CP(R[citing pub] > 0) | CP(R[citing pub] = 0) | CP(R[cited pub] > 0) | CP(R[cited pub] = 0) | Title | Journal | First author | Year |
| **2519** | 2086 | 433 | 118 | 2401 | An index to quantify an individual's scientific research output | Proceedings of the National Academy of Sciences of the United States of America | HIRSCH, JE | 2005 |
| **659** | 450 | 209 | 207 | 452 | The increasing dominance of teams in production of knowledge | Science | WUCHTY, S | 2007 |
| **582** | 483 | 99 | 535 | 47 | Theory and practise of the g-index | Scientometrics | EGGHE, L | 2006 |
| **481** | 279 | 202 | 110 | 371 | The scientific impact of nations | Nature | KING, DA | 2004 |
| **462** | 262 | 200 | 292 | 170 | Coauthorship networks and patterns of scientific collaboration | Proceedings of the National Academy of Sciences of the United States of America | NEWMAN, MEJ | 2004 |
| **450** | 224 | 226 | 17 | 433 | Comparison of PubMed, Scopus, Web of Science, and Google Scholar: Strengths and weaknesses | FASEB Journal | FALAGAS, ME | 2008 |
| **400** | 188 | 212 | 21 | 379 | A guide for naming research studies in Psychology | International Journal of Clinical and Health Psychology | MONTERO, I | 2007 |
| **397** | 214 | 183 | 197 | 200 | Science faculty's subtle gender biases favor male students | Proceedings of the National Academy of Sciences of the United States of America | MOSS-RACUSIN, CA | 2012 |
| **375** | 267 | 108 | 242 | 133 | The impact of research collaboration on scientific productivity | Social Studies of Science | LEE, S | 2005 |
| **355** | 245 | 110 | 229 | 126 | Impact of data sources on citation counts and rankings of LIS faculty: Web of Science versus Scopus and Google Scholar | Journal of the American Society for Information Science and Technology | MEHO, LI | 2007 |
| Ranked by CP(R[citing pub] > 0) | | | | | | | | |
| CP | CP(R[citing pub] > 0) | CP(R[citing pub] = 0) | CP(R[cited pub] > 0) | CP(R[cited pub] = 0) | Title | Journal | First author | Year |
| 2519 | **2086** | 433 | 118 | 2401 | An index to quantify an individual's scientific research output | Proceedings of the National Academy of Sciences of the United States of America | HIRSCH, JE | 2005 |

| CP | CP(R[citing pub] > 0) | CP(R[citing pub] = 0) | CP(R[cited pub] > 0) | CP(R[cited pub] = 0) | Title | Journal | First author | Year |
|---|---|---|---|---|---|---|---|---|
| 582 | **483** | 99 | 535 | 47 | Theory and practise of the g-index | Scientometrics | EGGHE, L | 2006 |
| 659 | **450** | 209 | 207 | 452 | The increasing dominance of teams in production of knowledge | Science | WUCHTY, S | 2007 |
| 481 | **279** | 202 | 110 | 371 | The scientific impact of nations | Nature | KING, DA | 2004 |
| 375 | **267** | 108 | 242 | 133 | The impact of research collaboration on scientific productivity | Social Studies of Science | LEE, S | 2005 |
| 462 | **262** | 200 | 292 | 170 | Coauthorship networks and patterns of scientific collaboration | Proceedings of the National Academy of Sciences of the United States of America | NEWMAN, MEJ | 2004 |
| 355 | **245** | 110 | 229 | 126 | Impact of data sources on citation counts and rankings of LIS faculty: Web of Science versus Scopus and Google Scholar | Journal of the American Society for Information Science and Technology | MEHO, LI | 2007 |
| 327 | **240** | 87 | 287 | 40 | Does the h-index have predictive power? | Proceedings of the National Academy of Sciences of the United States of America | HIRSCH, JE | 2007 |
| 286 | **234** | 52 | 249 | 37 | Comparison of the Hirsch-index with standard bibliometric indicators and with peer judgment for 147 chemistry research groups | Scientometrics | VAN RAAN, AFJ | 2006 |
| 450 | **224** | 226 | 17 | 433 | Comparison of PubMed, Scopus, Web of Science, and Google Scholar: Strengths and weaknesses | FASEB Journal | FALAGAS, ME | 2008 |
| Ranked by CP(R[citing pub] = 0) | | | | | | | | |
| CP | CP(R[citing pub] > 0) | CP(R[citing pub] = 0) | CP(R[cited pub] > 0) | CP(R[cited pub] = 0) | Title | Journal | First author | Year |
| 2519 | 2086 | **433** | 118 | 2401 | An index to quantify an individual's scientific research output | Proceedings of the National Academy of Sciences of the United States of America | HIRSCH, JE | 2005 |
| 450 | 224 | **226** | 17 | 433 | Comparison of PubMed, Scopus, Web of Science, and Google Scholar: Strengths and weaknesses | FASEB Journal | FALAGAS, ME | 2008 |
| 400 | 188 | **212** | 21 | 379 | A guide for naming research studies in Psychology | International Journal of Clinical and Health Psychology | MONTERO, I | 2007 |
| 659 | 450 | **209** | 207 | 452 | The increasing dominance of teams in production of knowledge | Science | WUCHTY, S | 2007 |
| 481 | 279 | **202** | 110 | 371 | The scientific impact of nations | Nature | KING, DA | 2004 |
| 462 | 262 | **200** | 292 | 170 | Coauthorship networks and patterns of scientific collaboration | Proceedings of the National Academy of Sciences of the United States of America | NEWMAN, MEJ | 2004 |
| 253 | 69 | **184** | 0 | 253 | Who's afraid of peer review? | Science | BOHANNON, J | 2013 |

| CP | CP(R[citing pub] > 0) | CP(R[citing pub] = 0) | CP(R[cited pub] > 0) | CP(R[cited pub] = 0) | Title | Journal | First author | Year |
|---|---|---|---|---|---|---|---|---|
| 397 | 214 | **183** | 197 | 200 | Science faculty's subtle gender biases favor male students | Proceedings of the National Academy of Sciences of the United States of America | MOSS-RACUSIN, CA | 2012 |
| 186 | 39 | **147** | 33 | 153 | The rate of growth in scientific publication and the decline in coverage provided by Science Citation Index | Scientometrics | LARSEN, PO | 2010 |
| 345 | 200 | **145** | 189 | 156 | What do citation counts measure? A review of studies on citing behavior | Journal of Documentation | BORNMANN, L | 2008 |
| Ranked by CP(R[cited pub] > 0) | | | | | | | | |
| CP | CP(R[citing pub] > 0) | CP(R[citing pub] = 0) | CP(R[cited pub] > 0) | CP(R[cited pub] = 0) | Title | Journal | First author | Year |
| 582 | 483 | 99 | **535** | 47 | Theory and practise of the g-index | Scientometrics | EGGHE, L | 2006 |
| 462 | 262 | 200 | **292** | 170 | Coauthorship networks and patterns of scientific collaboration | Proceedings of the National Academy of Sciences of the United States of America | NEWMAN, MEJ | 2004 |
| 327 | 240 | 87 | **287** | 40 | Does the h-index have predictive power? | Proceedings of the National Academy of Sciences of the United States of America | HIRSCH, JE | 2007 |
| 286 | 234 | 52 | **249** | 37 | Comparison of the Hirsch-index with standard bibliometric indicators and with peer judgment for 147 chemistry research groups | Scientometrics | VAN RAAN, AFJ | 2006 |
| 375 | 267 | 108 | **242** | 133 | The impact of research collaboration on scientific productivity | Social Studies of Science | LEE, S | 2005 |
| 241 | 204 | 37 | **236** | 5 | The R- and AR-indices: Complementing the h-index | Chinese Science Bulletin | JIN, BH | 2007 |
| 355 | 245 | 110 | **229** | 126 | Impact of data sources on citation counts and rankings of LIS faculty: Web of Science versus Scopus and Google Scholar | Journal of the American Society for Information Science and Technology | MEHO, LI | 2007 |
| 276 | 209 | 67 | **223** | 53 | Universality of citation distributions: Toward an objective measure of scientific impact | Proceedings of the National Academy of Sciences of the United States of America | RADICCHI, F | 2008 |
| 659 | 450 | 209 | **207** | 452 | The increasing dominance of teams in production of knowledge | Science | WUCHTY, S | 2007 |
| 222 | 188 | 34 | **206** | 16 | Is it possible to compare researchers with different scientific interests? | Scientometrics | BATISTA, PD | 2006 |
| Ranked by CP(R[cited pub] = 0) | | | | | | | | |
| CP | CP(R[citing pub] > 0) | CP(R[citing pub] = 0) | CP(R[cited pub] > 0) | CP(R[cited pub] = 0) | Title | Journal | First author | Year |
| 2519 | 2086 | 433 | 118 | **2401** | An index to quantify an individual's scientific research output | Proceedings of the National Academy of Sciences of the United States of America | HIRSCH, JE | 2005 |
| 659 | 450 | 209 | 207 | **452** | The increasing dominance of teams in production of knowledge | Science | WUCHTY, S | 2007 |
| 450 | 224 | 226 | 17 | **433** | Comparison of PubMed, Scopus, Web of Science, and Google Scholar: Strengths and weaknesses | FASEB Journal | FALAGAS, ME | 2008 |
| 400 | 188 | 212 | 21 | **379** | A guide for naming research studies in Psychology | International Journal of Clinical and Health Psychology | MONTERO, I | 2007 |
| 481 | 279 | 202 | 110 | **371** | The scientific impact of nations | Nature | KING, DA | 2004 |
| 288 | 185 | 103 | 0 | **288** | Free online availability substantially increases a paper's impact | Nature | LAWRENCE, S | 2001 |
| 253 | 69 | 184 | 0 | **253** | Who's afraid of peer review? | Science | BOHANNON, J | 2013 |
| 266 | 169 | 97 | 49 | **217** | Journal prestige, publication bias, and other characteristics associated with citation of published studies in peer-reviewed journals | Journal of the American Medical Association | CALLAHAM, M | 2002 |
| 397 | 214 | 183 | 197 | **200** | Science faculty's subtle gender biases favor male students | Proceedings of the National Academy of Sciences of the United States of America | MOSS-RACUSIN, CA | 2012 |
| 315 | 215 | 100 | 121 | **194** | Rankings and reactivity: How public measures recreate social worlds | American Journal of Sociology | ESPELAND, WN | 2007 |